\documentclass{article}

\usepackage[final]{neurips_2019}

\usepackage[utf8]{inputenc} 
\usepackage[T1]{fontenc}    
\usepackage{hyperref}       
\usepackage{url}            
\usepackage{booktabs}       
\usepackage{amsfonts}       
\usepackage{nicefrac}       
\usepackage{microtype}      

\setcitestyle{numbers}
\setcitestyle{square}
\setcitestyle{comma}
\usepackage{graphicx}

\title{Leveraging Topics and Audio Features with Multimodal Attention for Audio~Visual~Scene-Aware~Dialog}

\author{
  Shachi H Kumar \\
  Intel Labs \\
  Anticipatory Computing Lab \\
  Santa Clara, CA 95054 \\
  \texttt{shachi.h.kumar@intel.com} \\
  \And
  Eda Okur \\
  Intel Labs \\
  Anticipatory Computing Lab \\
  Hillsboro, OR 97124 \\
  \texttt{eda.okur@intel.com} \\
  \And
  Saurav Sahay \\
  Intel Labs \\
  Anticipatory Computing Lab \\
  Santa Clara, CA 95054 \\
  \texttt{saurav.sahay@intel.com} \\
  \And
  Jonathan Huang \\
  Intel Labs \\
  Anticipatory Computing Lab \\
  Santa Clara, CA 95054 \\
  \texttt{jonathan.huang@intel.com} \\
  \And
  Lama Nachman \\
  Intel Labs \\
  Anticipatory Computing Lab \\
  Santa Clara, CA 95054 \\
  \texttt{lama.nachman@intel.com} \\
}

\begin{document}

\maketitle

\begin{abstract}
  With the recent advancements in Artificial Intelligence (AI), Intelligent Virtual Assistants (IVA) such as Alexa, Google Home, etc., have become a ubiquitous part of many homes. Currently, such IVAs are mostly audio-based, but going forward, we are witnessing a confluence of vision, speech and dialog system technologies that are enabling the IVAs to learn audio-visual groundings of utterances. This will enable agents to have conversations with users about the objects, activities and events surrounding them. In this work, we present three main architectural explorations for the Audio Visual Scene-Aware Dialog (AVSD): 1) investigating `topics' of the dialog as an important contextual feature for the conversation, 2) exploring several multimodal attention mechanisms during response generation, 3) incorporating an end-to-end audio classification ConvNet, AclNet, into our architecture. 
  We discuss detailed analysis of the experimental results and show that our model variations outperform the baseline system presented for the AVSD task. 
\end{abstract}

\section{Introduction}
We are witnessing a confluence of vision, speech and dialog system technologies that are enabling the IVAs to learn audio-visual groundings of utterances and have conversations with users about the objects, activities and events surrounding them. Recent progress in visual grounding techniques \cite{7410636, Das_2017} and audio understanding \cite{hershey2017cnn} are enabling machines to understand shared semantic concepts and listen to the various sensory events in the environment. With audio and visual grounding methods \cite{yu2016video, Hori_2017}, end-to-end multimodal Spoken Dialog Systems (SDS) \cite{serban2016building} are now being trained to meaningfully communicate in natural language about the real dynamic audio-visual sensory world around us. In this work, we explore the role of `topics' of the dialog as the context of the conversation along with multimodal attention into an end-to-end audio-visual scene-aware dialog system architecture. We also incorporate an end-to-end audio classification ConvNet, AclNet, into our models. We develop and test our approaches on the Audio Visual Scene-Aware Dialog (AVSD) dataset \cite{DBLP:journals/corr/abs-1806-00525, alamri2019audiovisual} released as part of the 7th Dialog System Technology Challenges (DSTC7) task, showing that some of our model variations outperform the AVSD baseline model \cite{DBLP:journals/corr/abs-1806-08409}.

\section{Model Description}
\begin{figure}
  \centering
  \includegraphics[width=\linewidth]{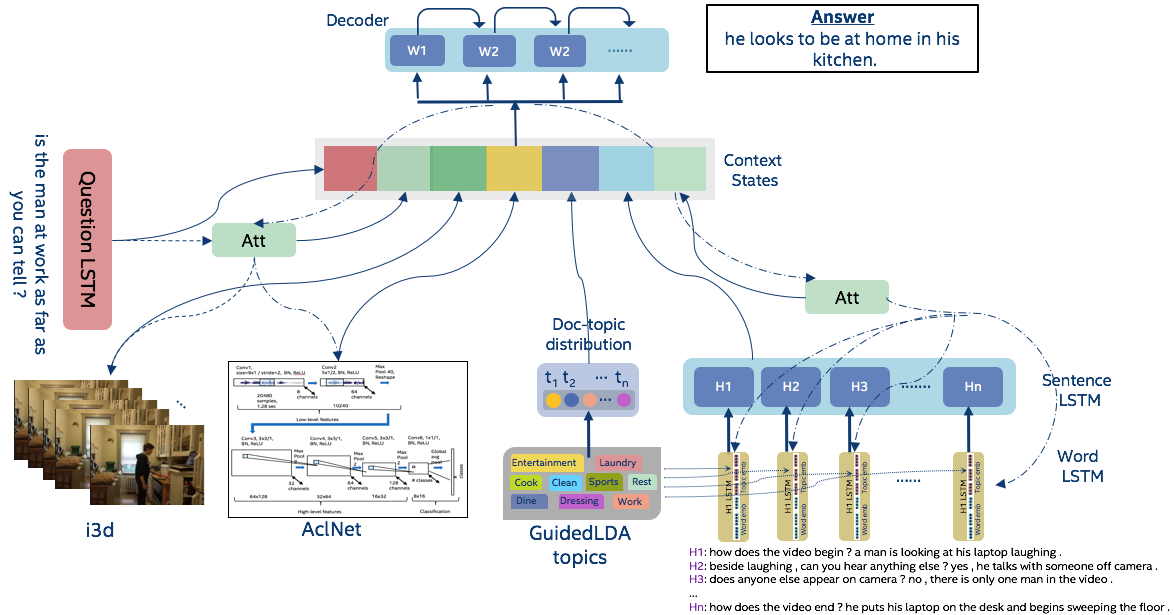}
  \caption{Architecture of Our System}
  \label{arch}
\end{figure}

In this section, we describe the main architectural explorations of our work as shown in Figure~\ref{arch}. 

\textbf{Adding Topics of Conversations}: 
Topics form a very important source of context in a dialog. 
Charades dataset \cite{DBLP:conf/eccv/SigurdssonVWFLG16} contains videos on common household activities such as watching TV, eating, cleaning, using a laptop, sleeping, and so on. 
We train Latent Dirichlet Allocation (LDA) \cite{DBLP:conf/nips/BleiNJ01} and Guided LDA 
\cite{DBLP:conf/eacl/JagarlamudiDU12} models on questions, answers, QA pairs, captions and dialog history. 
Since we are interested in identifying domain-specific topics such as entertainment, cooking, cleaning, resting, etc., we use Guided LDA to generate topics via seed words. A detailed list of sample seed words provided to Guided LDA for the 9-topics configuration is presented in Table~\ref{table:seeds9}. These seed words are constructed by identifying a set of most common nouns (objects), verbs, scenes, and actions from the Charades dataset analysis \cite{DBLP:conf/eccv/SigurdssonVWFLG16}. Generated topic distributions are incorporated as features into our models or used to learn topic embeddings.

\begin{table}[!b]
  \caption{Sample of Seed Words for 9 Topics}   
  \label{table:seeds9}
  \centering
  \scriptsize 
  \resizebox{\columnwidth}{!}{
  \begin{tabular}{ l  l }
    \toprule
    \textbf{Topic} & \textbf{Seed Words} \\ 
    \midrule
  Entertainment/LivingRoom & \tiny living, room, recreation, garage, basement, entryway, television, tv, phone, laptop, sofa, chair, couch, armchair, seat, picture, sit ... \\ 
    \midrule
  Cooking/Kitchen & \tiny kitchen, pantry, food, water, dish, sink, refrigerator, fridge, stove, microwave, toaster, kettle, oven, stewpot, saucepan, cook, wash ... \\ 
    \midrule
  Eating/Dining & \tiny dining, room, table, chair, plate, fork, knife, spoon, bowl, glass, cup, mug, coffee, tea, sandwich, meal, breakfast, lunch, dinner ... \\ 
    \midrule
  Cleaning/Bath & \tiny bathroom, hallway, entryway, stairs, restroom, toilet, towel, broom, vacuum, floor, sink, water, mirror, cabinet, hairdryer, clean ... \\ 
    \midrule
  Dressing/Closet & \tiny walk-in, closet, clothes, wardrobe, shoes, shirt, pants, trousers, skirt, jacket, t-shirt, underwear, sweatshirt, coat, rack, dress, wear ... \\ 
    \midrule
  Laundry & \tiny laundry, room, basement, clothes, clothing, cloth, basket, bag, box, towel, shelf, dryer, washer, washing, machine, do, wash, hold ... \\ 
    \midrule
  Rest/Bedroom & \tiny bedroom, room, bed, pillow, blanket, mattress, bedstand, nightstand, commode, dresser, bedside, lamp, nightlight, night, light, lie ... \\ 
    \midrule
  Work/Study & \tiny home, office, den, workroom, garage, basement, laptop, computer, pc, screen, mouse, keyboard, phone, desk, chair, light, work, study ... \\ 
    \midrule
  Sports/Exercise & \tiny recreation, room, garage, basement, hallway, stairs, gym, fitness, floor, bag, towel, ball, treadmill, bike, rope, mat, run, walk, exercise ... \\ 
    \bottomrule
  \end{tabular}
  }
\end{table}

\textbf{Attention Explorations}:  
 We explore several configurations of the attention-based model where at every step, the decoder attends to the dialog history representations and audio/video (AV) features to selectively focus on relevant parts of the dialog history and AV. We calculate the attention weights \cite{DBLP:journals/corr/BahdanauCB14, DBLP:conf/nips/VaswaniSPUJGKP17} corresponding to every dialog history turn, multimodal features and the decoder representation, and apply the weights to the history and multimodal features to compute the relevant representations. 
 These help create a combination of the dialog history and multimodal context that is richer than the single context vectors of the individual modalities. We append the input encoding along with the AV multimodal feature encodings and pass that to the decoder LSTM for learning the output encodings.
 
\textbf{Audio Feature Explorations}:
We used an end-to-end audio classification ConvNet, called AclNet \cite{AclNet}. AclNet takes raw, amplitude-normalized 44.1 kHz audio samples as input, and produces classification output without the need to compute spectral features. AclNet is trained using the ESC-50 \cite{piczak2015dataset} corpus, a dataset of 50 classes of environmental sounds organized in 5 semantic categories (animals, interior/domestic, exterior/urban, human, natural landscapes).

\section{Dataset}
We use the dialog dataset consisting of conversations between two parties about short videos (from Charades human action dataset~\cite{DBLP:conf/eccv/SigurdssonVWFLG16}), which was released as part of the AVSD challenge track of DSTC7 \cite{DBLP:journals/corr/abs-1806-00525}. The two parties in the conversation discuss about events in the video, where one plays the role of a questioner and the other is the answerer \cite{alamri2019audiovisual}. For the results presented in this work, we use the official training and validation sets to train and optimize our models, which are evaluated on the official test set. Table~\ref{table:dataset} shows the distribution of DSTC7 AVSD data across different sets. Further details of our AVSD dataset analysis and previous results on prototype sets can be found in \cite{kumar2019context}. 

\begin{table}[!h]
  \caption{Audio Visual Scene-Aware Dialog Dataset}
  \label{table:dataset}
  \centering
  \small
  \begin{tabular}{lccc}
    \toprule
    & Training & Validation & Test \\ 
    \midrule
    \# of Dialogs & 7,659 & 1,787 & 1,710 \\
    \# of Turns & 153,180 & 35,740 & 13,490 \\
    \# of Words & 1,450,754 & 339,006 & 110,252 \\
    \bottomrule
  \end{tabular}
\end{table}

\section{Experiments and Results}

\begin{table}[!b]
  \caption{Topic Modeling Experiments}
  \label{table:all-topics}
  \centering
  \scriptsize
  \begin{tabular}{ l  c  c  c  c  c  c  c }
    \toprule
    & BLEU$_1$ &  BLEU$_2$ & BLEU$_3$ & BLEU$_4$ & METEOR & ROUGE$_L$ & CIDEr \\
    \midrule
    Baseline & 
    0.621 & 0.480 & 0.379 & 0.305 & 0.217 & 0.481 & 0.733 \\
    \hline
    GuidedLDA (Q,QA,C) & 	
    0.614 & 0.475 & 0.374 & 0.299 & 0.215 & 0.474 & 0.695 \\		
    GuidedLDA (Q,QA,C) + GloVe 
    & 0.629	& 0.491	& 0.390	& 0.315	& 0.219	& 0.484	& 0.731	\\		
    StandardLDA (All topics) 
    & 0.621	& 0.480	& 0.380	& 0.306	& 0.221	& 0.483 & 0.753\\
    GuidedLDA (All topics) 
    & 0.619	& 0.480	& 0.378	& 0.303	& 0.217	& 0.476	& 0.701	\\		
    GuidedLDA (All topics) + GloVe 
    & \textbf{0.631} & \textbf{0.493} & \textbf{0.390} & \textbf{0.315} & \textbf{0.224}	& \textbf{0.492} & \textbf{0.773} \\				
    \hline
    {HLSTM with topics}	
    & \textbf{0.627} & \textbf{0.489} & \textbf{0.387} & \textbf{0.311} & \textbf{0.218}	& 0.480 & 0.723	\\				\hline
    Topic Embeddings 
    & 0.623	& 0.488	& 0.387	& 0.311	& 0.217	& 0.479	& 0.701	\\		
    Topic Embeddings + GloVe 
    & \textbf{0.632} & \textbf{0.499} & \textbf{0.402} & \textbf{0.329} & \textbf{0.223}	& \textbf{0.488} & \textbf{0.762} \\												
    \bottomrule
  \end{tabular}
\end{table}

\begin{table}[!b]
  \caption{Topic Model Performances on Binary/Non-binary Answers}
  \label{table:topicBinNonBin}
  \centering
  \scriptsize
  \begin{tabular}{ l  c  c  c  c  c  c  c }
    \toprule
    & BLEU$_1$ &  BLEU$_2$ & BLEU$_3$ & BLEU$_4$ & METEOR & ROUGE$_L$ & CIDEr \\
    \midrule
    \textbf{Binary} & & & & & & & \\
Baseline
&  0.626
&  0.479
&  0.371
&  0.294
&  0.214
&  0.474
&  0.676
\\ 
GuidedLDA (Q,QA,C) + GloVe
&   0.616
&   0.476
&   0.374
&   0.301
&   0.215
&   0.474
&   0.673
\\
GuidedLDA (All topics) + GloVe
&   0.629
&   0.486
&   0.381
&   0.306
&   0.223
&   0.488
&   0.728
\\
HLSTM with topics
&   0.623
&   0.480
&   0.375
&   0.297
&   0.214
&   0.473
&   0.696
\\
Topic Embeddings + GloVe
& \textbf{0.635}
& \textbf{0.497}
& \textbf{0.398}
& \textbf{0.325}
& \textbf{0.224}
& \textbf{0.491}
& \textbf{0.746}
\\
    \midrule
    \textbf{Non-binary} & & & & & & & \\
Baseline
&  0.624
&  0.486
&  0.387
&  0.312
&  0.219
&  0.482
&  0.726
\\
GuidedLDA (Q,QA,C) + GloVe
&  \textbf{0.633}
&  0.497
&  0.396
&  0.320
&  0.220
&  0.487
&  0.759
\\
GuidedLDA (All topics) + GloVe
&  0.632
&  0.495
&  0.394
&  0.318
&  \textbf{0.225}
&  \textbf{ 0.494}
&  \textbf{0.796}
\\
HLSTM with topics
&  0.629
&  0.492
&  0.392
&  0.316
&  0.220
&  0.483
&  0.740
\\
Topic Embeddings + GloVe
&  0.630
&  \textbf{0.499}
&  \textbf{0.403}
&  \textbf{0.330}
&  0.223
&  0.487
&  0.776 
\\
    \bottomrule
    \end{tabular}
\end{table}

\textbf{Topic Modeling Experiments}:
We use separate topic models trained on questions (Q), answers (A), QA pairs, captions (C), history and history+captions to generate topics for samples from each category. The generated topic vectors are incorporated as features for questions and dialog history. The question topics are added to the decoder state directly. In one variation, the dialog history topics (QA and C, or all topics) are copied to all the decoder states directly. In another variation, the dialog history topics are added as features to the history encoder LSTM (HLSTM). We learn topic embeddings from topics generated for the questions, QA pairs and captions as well. In addition, GloVe embeddings \cite{pennington2014glove} (200-dim) are incorporated with fine-tuning for questions and history.

Table~\ref{table:all-topics} compares the baseline model \cite{DBLP:journals/corr/abs-1806-08409} with the topic-based model variations. GuidedLDA (All topics) + GloVe performs better than the baseline in all metrics. Adding topics as part of the HLSTM also slightly improves performance compared to the baseline. Learning topic embeddings along with the word embeddings (+GloVe fine-tuning) achieves the best performance in most of the metrics (BLEU-scores), whereas GuidedLDA (All topics) + GloVe succeeds in other metrics. We also evaluated topic-based models on subsets having binary and non-binary answers. As shown in Table~\ref{table:topicBinNonBin}, for the non-binary subset, all topic-based models perform better than the baseline in all metrics, which shows that these models can generate better responses for the more complex, non-binary answers.

\begin{figure}[!h]
  \centering
  \includegraphics[width=\textwidth]{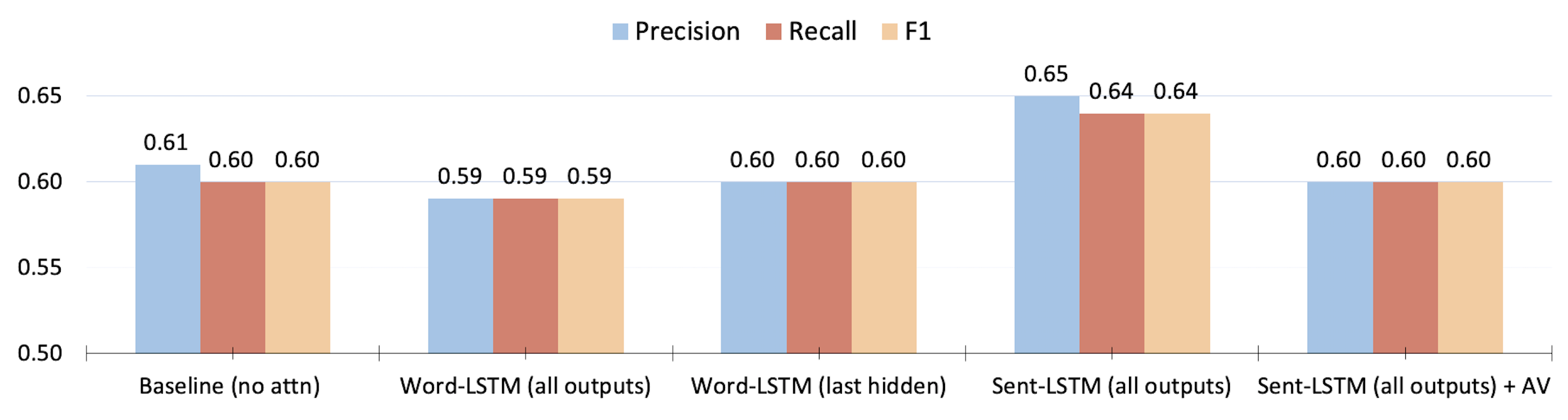}
  \caption{Precision, Recall, F1-scores for Attention Experiments on Coreference-Binary-subset}
  \label{attentionCoref}
\end{figure}

\textbf{Attention Experiments}: 
The baseline architecture \cite{DBLP:journals/corr/abs-1806-08409} only leverages the last hidden state information from the sentence LSTM in the dialog history encoder. 
In our experiments, we have modified the baseline architecture and added attention layer for the answer decoder to leverage information directly from the dialog history LSTMs and multimodal audio/video features, with 4 different configurations described below. 
To evaluate the performance of attention solely for questions that could benefit from dialog history, we isolate the questions containing coreferences. Table~\ref{table:attn} shows the performance of our models on this coreference-subset. To compare the results at a more semantic level, we further performed quantitative analysis on dialogs that contained binary answers. We evaluate our models on their ability to predict these binary answers correctly (using precision, recall and F1-scores) as presented in Figure~\ref{attentionCoref}. The results show that the configuration where decoder attends to all of the sentence-LSTM output states performs better than the baseline. 

\begin{table}[!h]
  \caption{Decoder Attention over Dialog History and Multimodal Features on Coreference-subset}    
  \label{table:attn}
  \centering
  \scriptsize
  \begin{tabular}{ l  c  c  c  c  c  c  c }
    \toprule
    & BLEU$_1$ &  BLEU$_2$ & BLEU$_3$ & BLEU$_4$ & METEOR & ROUGE$_L$ & CIDEr \\
    \midrule
    Baseline & 0.611 & 0.475 & 0.374 & 0.297 & 0.210 & 0.467 & 0.704 \\ 
    Word LSTM (all output states) & 0.594 & 0.447 & 0.336 & 0.262 & 0.190 & 0.432 & 0.553  \\ 
    Word LSTM (last hidden states) & \textbf{0.627} & \textbf{0.485} & \textbf{0.379} & 0.297 & 0.208 & \textbf{0.468} & 0.701 \\ 
    Sentence LSTM (all output states) & \textbf{0.619} & \textbf{0.484} & \textbf{0.384} & \textbf{0.307} & \textbf{0.213} & \textbf{0.472} & \textbf{0.749} \\
    Sentence LSTM (all outputs) + AV & 0.598 & 0.464 & 0.360 & 0.284 & 0.209 & 0.458 & 0.685 \\
    \bottomrule
  \end{tabular}
\end{table}

\begin{itemize}
\item[1.] \emph{Attention on Dialog History Word LSTMs, all output states}: In this configuration, we remove the sentence level dialog history LSTM and the decoder computes the attention scores directly between the decoder state and the word level output states for all dialog history. We first padded the Word LSTM outputs from Dialog History LSTMs (see Word LSTM in History in Figure~\ref{arch}) to the maximum sentence length of all the sentences. We summed up all the attention scores from each of the sentence context vectors with the query decoder state. Using this kind of attention, we had hoped that the system could remember answers that were already given (directly or indirectly) in the earlier turns of the dialog. Directly attending to the output states of the word LSTMs in the dialog history encoder did not perform well compared to the baseline. This attention mechanism possibly attended to way more information than needed.
\item[2.] \emph{Attention on Dialog History Word LSTMs, last hidden states}: This configuration is similar to the previous configuration with the difference that we only use the last hidden state output representations of the word LSTMs corresponding to the different turns in the dialog. Simpler than the previous setup, we stack up the hidden states from the history sentences for attention computation. This configuration performed better that the baseline on the coreference-subset in most of the evaluation metrics.
\item[3.] \emph{Attention on Sentence LSTM, all output states}: The baseline architecture only leverages the last hidden state information from the sentence LSTM in the dialog history encoder. Instead, we extract the output states from all timesteps of the LSTM corresponding to $n$ turns of the dialog history. This variation helps the decoder consider all the dialog turn compressed sentence representations via the attention mechanism. This model performed better than the baseline in all metrics on both coreference-subset (Table~\ref{table:attn}) and binary answers (Figure~\ref{attentionCoref}).
\item[4.] \emph{Attention on Sentence LSTM, all output states and Multimodal Audio/Video Features}: This configuration is similar to the last one with the difference that we add multimodal audio/video features as additional state to the attention module. This mechanism allows the decoder to selectively focus on the multimodal features along with the dialog history sentences. This configuration did not really help improve the evaluation metrics compared to the baseline.
\end{itemize}

\textbf{Audio Experiments}:
Table~\ref{table:audioexp_final} shows the comparison of the baseline (B) model without audio features, B+VGGish (provided as a part of the AVSD task), and B+AclNet features. We investigate the effects of audio features on the overall dataset as well as on the subset of audio-related questions. We observe that B+AclNet shows improved performances as compared to the baseline and B+VGGish, both on the overall dataset and audio-related subset. Table~\ref{table:qualitativeaudio} presents a qualitative analysis of the addition of the VGGish and AclNet features to the baseline model. For these audio-related examples (e.g., 'oscillating', 'eating', 'sneeze'), baseline and B+VGGish models generate irrelevant responses, whereas the answers generated by B+AclNet are in accordance with the ground truth.

\begin{table}[!h]
  \caption{Audio Feature Performances on Overall vs. Audio-related Questions} 
  \label{table:audioexp_final}
  \centering
  \scriptsize
  \begin{tabular}{ l  c  c  c  c  c  c  c }
    \toprule
    & BLEU$_1$ &  BLEU$_2$ & BLEU$_3$ & BLEU$_4$ & METEOR & ROUGE$_L$ & CIDEr \\
    \midrule
    \textbf{Overall} & & & & & & & \\
    Baseline (B) 
    & 0.621 & 0.480 & 0.379 & 0.305 & 0.217 & 0.481 & 0.733 \\
    B + VGGish 
    & 0.622 & 0.487 & 0.389 & 0.315 & 0.216 & 0.481 & 0.732 \\
    B + AclNet 
    & \textbf{0.625} & \textbf{0.491} & \textbf{0.391} & \textbf{0.316} & \textbf{0.218} & \textbf{0.484} & \textbf{0.736}\\
    \midrule
    \textbf{Audio-related} & & & & & & & \\
    Baseline (B) 
    &\textbf{0.666} & 0.526 & 0.413 & 0.329 & 0.230 & 0.504 & 0.767 \\
    B + VGGish 
    & 0.657 & 0.519 & 0.408 & 0.324 & 0.230 & 0.500 &  0.754 \\ 
    B + AclNet 
    & 0.659 & \textbf{0.527} & \textbf{0.424} & \textbf{0.348} & \textbf{0.236} & \textbf{0.507} & \textbf{0.796} \\ 
    \bottomrule
  \end{tabular}
\end{table}

\begin{table*}[!h]
  \caption{Audio Examples (VGGish vs. AclNet)}    \label{table:qualitativeaudio}
  \centering
  \scriptsize
  \begin{tabular}{ l  l  l  l  l }
    \toprule
    \textit{Question:} & \textit{\textbf{is the fan oscillating ?}} & \textit{\textbf{is he eating something ?}} & \textit{\textbf{how many times does she sneeze ? }} \\     
    \midrule
    Ground Truth & \textit{the fan is on but is still .} & \textit{yes he appears to be eating something} & \textit{she sneezes a few times in the video .} \\
    \midrule
    Baseline & \textit{yes it is very well lit} & \textit{no he is not drinking anything} & \textit{can only see her face} \\     
    \midrule
    Baseline + VGGish & \textit{no don 't see any music}	& \textit{no he is not drinking anything} & \textit{she laughs at the end of the video} \\
    \midrule
    Baseline + AclNet & \textit{no it is hard to tell} & \textit{yes he is eating sandwich} & \textit{she sneezes at the end of the video} \\
    \bottomrule
  \end{tabular}
\end{table*}

\section{Conclusion}
In this paper, we present our explorations towards architectural extensions for contextual and multimodal end-to-end audio-visual scene-aware dialog system. We incorporate context of the dialog in the form of topics, investigate various attention mechanisms to enable the decoder to focus on relevant parts of the dialog history and audio/video features, and incorporate audio features from an end-to-end audio classification architecture, AclNet. We validate our approaches on the AVSD dataset and show that some of the explored techniques yields in improved performances compared to the baseline system for AVSD task.

\small

\bibliography{ViGIL_cameraReady_arXiv}

\end{document}